\documentclass[
aps,%
10pt,%
final,%
notitlepage,%
oneside,%
twocolumn,%
nobibnotes,%
nofootinbib,
superscriptaddress,%
noshowpacs,%
]%
{revtex4}

\usepackage{graphicx}

\begin{document}

\title{\boldmath $a_0^+(980)$-resonance production in the reaction
 $pp \to d\pi^+\eta$ close to the $K\bar{K}$ threshold}

\author{P.~Fedorets}
\affiliation{Institut f\"ur Kernphysik, Forschungszentrum J\"ulich, 52425 J\"ulich, Germany}
\affiliation{Institute for Theoretical and Experimental Physics,
  Bolshaya Cheremushkinskaya 25, 117218 Moscow, Russia}

\author{M.~B\"uscher}
\affiliation{Institut f\"ur Kernphysik, Forschungszentrum J\"ulich, 52425 J\"ulich, Germany}

\author{V.P.~Chernyshev}
\affiliation{Institute for Theoretical and Experimental Physics, Bolshaya Cheremushkinskaya 25, 117218 Moscow, Russia}

\author{S.~Dymov}
\affiliation{Institut f\"ur Kernphysik, Forschungszentrum J\"ulich, 52425 J\"ulich, Germany}
\affiliation{Laboratory of Nuclear Problems, Joint Institute for Nuclear Research, 141980 Dubna, Russia}

\author{V.Yu.~Grishina}
\affiliation{Institute for Nuclear Research, 60th October Anniversary Prospect 7A, 117312 Moscow, Russia}

\author{C.~Hanhart}
\affiliation{Institut f\"ur Kernphysik, Forschungszentrum J\"ulich, 52425 J\"ulich, Germany}

\author{M.~Hartmann}
\affiliation{Institut f\"ur Kernphysik, Forschungszentrum J\"ulich, 52425 J\"ulich, Germany}

\author{V.~Hejny}
\affiliation{Institut f\"ur Kernphysik, Forschungszentrum J\"ulich, 52425 J\"ulich, Germany}

\author{V.~Kleber}
\affiliation{Institut f\"ur Kernphysik, Universit\"at zu K\"oln, Z\"ulpicher Str.\ 77, 50937 K\"oln, Germany}

\author{H.R.~Koch}
\affiliation{Institut f\"ur Kernphysik, Forschungszentrum J\"ulich, 52425 J\"ulich, Germany}

\author{L.A.~Kondratyuk}
\affiliation{Institute for Theoretical and Experimental Physics, Bolshaya Cheremushkinskaya 25, 117218 Moscow, Russia}

\author{V.~Koptev}
\affiliation{High Energy Physics Department, Petersburg Nuclear Physics Institute, 188350 Gatchina, Russia}

\author{A.E.~Kudryavtsev}
\affiliation{Institute for Theoretical and Experimental Physics, Bolshaya Cheremushkinskaya 25, 117218 Moscow, Russia}

\author{P.~Kulessa}
\affiliation{Institut f\"ur Kernphysik, Forschungszentrum J\"ulich, 52425 J\"ulich, Germany}

\author{S.~Merzliakov}
\affiliation{Laboratory of Nuclear Problems, Joint Institute for Nuclear Research, 141980 Dubna, Russia}

\author{S.~Mikirtychiants}
\affiliation{High Energy Physics Department, Petersburg Nuclear Physics Institute, 188350 Gatchina, Russia}

\author{M.~Nekipelov}
\affiliation{Institut f\"ur Kernphysik, Forschungszentrum J\"ulich, 52425 J\"ulich, Germany}
\affiliation{High Energy Physics Department, Petersburg Nuclear Physics Institute, 188350 Gatchina, Russia}

\author{H.~Ohm}
\affiliation{Institut f\"ur Kernphysik, Forschungszentrum J\"ulich, 52425 J\"ulich, Germany}

\author{R.~Schleichert}
\affiliation{Institut f\"ur Kernphysik, Forschungszentrum J\"ulich, 52425 J\"ulich, Germany}

\author{H.~Str\"oher}
\affiliation{Institut f\"ur Kernphysik, Forschungszentrum J\"ulich, 52425 J\"ulich, Germany}

\author{V.E.~Tarasov}
\affiliation{Institute for Theoretical and Experimental Physics, Bolshaya Cheremushkinskaya 25, 117218 Moscow, Russia}

\author{K.-H.~Watzlawik}
\affiliation{Institut f\"ur Kernphysik, Forschungszentrum J\"ulich, 52425 J\"ulich, Germany}

\author{I.~Zychor}
\affiliation{The Andrzej Soltan Institute for Nuclear Studies, 05400 Swierk, Poland}

\begin{abstract}
The reaction $pp\to d\pi^+\eta$ has been measured at a beam energy
of $T_p$=2.65~GeV ($p_p$=3.46~GeV/c) using the ANKE spectrometer
at COSY-J\"ulich. The missing mass distribution of the detected
$d\pi^+$ pairs exhibits a peak around the $\eta$ mass on top of a
strong background of multi-pion $pp\to d\pi^+ (n\pi)$ events. The
differential cross section
${\mathrm d}^4\sigma/{\mathrm d}{\Omega_d}{\mathrm d}
{\Omega_{\pi^+}}{\mathrm d}p_d{\mathrm d}p_{\pi^+}$
for the reaction $pp \to d\pi^+\eta$ has been determined model
independently for two regions of phase space. Employing a
dynamical model for the $a_0^+$ production allows one then to
deduce a total cross section of $\sigma(pp\to da_0^+ \to
d\pi^+\eta)=(1.1\pm 0.3_\mathrm{stat}\pm 0.7_\mathrm{sys})~\mu$b
for the production of $\pi^+\eta$ via the scalar $a_0^+$(980)
resonance and $\sigma(pp\to d\pi^+\eta) = (3.5\pm
0.3_{\mathrm{stat}}\pm 1.0_{\mathrm{sys}})~\mu$b for the
non-resonant production. Using the same model as for the
interpretation of recent results from ANKE for the reaction
$pp\to dK^+\bar{K^0}$, the ratio of the total cross sections is
$\sigma(pp\to d{(K^+\bar{K^0})}_{L=0})/\sigma(pp\to da_0^+\to
d\pi^+\eta) = 0.029 \pm 0.008_{\mathrm{stat}} \pm
0.009_{\mathrm{sys}}$, which is in agreement with branching ratios
in the literature.
\end{abstract}
\maketitle

\section{Introduction}
\vspace*{-4mm}

One of the primary goals of hadronic physics is the understanding
of the internal structure of mesons and baryons, and their
production and decay, in terms of quarks and gluons. The
non-perturbative character of the underlying theory --- Quantum
Chromo Dynamics (QCD) --- hinders straightforward calculations.
QCD can be treated explicitly in the low momentum-transfer regime
using lattice techniques~\cite{lattice}, which are, however, not
yet in the position to make quantitative statements about the
light scalar mesons. Alternatively, QCD-inspired models, which
employ effective degrees of freedom, can be used. The
constituent quark model is one of the most successful in this
respect (see e.g.~\cite{Morgan}). This approach treats the
lightest scalar resonances $a_0/f_0$(980) as conventional
$q\bar{q}$ states.

However, more states with quantum numbers $J^P{=}0^+$ have been
identified experimentally than would fit into a single SU(3)
scalar nonet: the $f_0(600)$ (or $\sigma$), $f_0(980)$,
$f_0(1370)$, $f_0(1500)$ and $f_0(1710)$ with $I{=}0$, the
$\kappa(800)$ and $K^*(1430)$ ($I{=}1/2$), as well as the
$a_0(980)$ and $a_0(1450)$ ($I{=}1$)~\cite{PDG}. Consequently, the
$a_0/f_0$(980) have also been associated with $K\bar{K}$
molecules~\cite{Wein} or compact $qq$-$\bar{q}\bar{q}$
states~\cite{Achasov}. It has even been suggested that a complete
nonet of four-quark states might exist with masses below
1.0~GeV/c$^2$~\cite{4q_nonet}.

The first clear observation of the isovector $a_0(980)$ resonance
was achieved in $K^-p$ interactions~\cite{Gay}, and in subsequent
experiments it has also been seen in $p\bar{p}$
annihilations~\cite{Abele}, in $\pi^-p$ collisions~\cite{Teig},
and in $\gamma\gamma$ interactions~\cite{Achard}. Experiments on
radiative $\phi$-decays~\cite{Achas,Alois} have been analysed in
terms of the $a_0/f_0$ production in the decay chain $\phi\to
\gamma{a_0}/ f_0\to \gamma\pi^0\eta/\pi^0\pi^0$. In $pp$
collisions the $a_0(980)$ resonance has been measured at
$p_p=450$~GeV/c \textit{via} $f_1(1285)\to a_0^{\pm}\pi^{\mp}$
decays~\cite{Barb} and in inclusive measurements of the $pp\to
dX^+$ reaction at $p_p=3.8$, 4.5, and 6.3~GeV/c~\cite{Abol}.
Despite these many experimental results, the properties of the
$a_0(980)$ resonance are still far from being established. The
Particle Data Group gives a mass of
$m_{a_0}=(984.7\pm1.2)$~MeV/c$^2$ and a width of
$\Gamma_{a_0}=(50-100)$~MeV/c$^2$~\cite{PDG}. The main decay
channels, $\pi\eta$ and $K\bar K$, are quoted as ``dominant'' and
``seen'' respectively.

An experimental programme has been started at the Cooler
Synchrotron COSY-J\"ulich~\cite{Maier} aiming at 
exclusive data on the $a_0/f_0$ production from $pp$, $pn$, $pd$
and $dd$ interactions at energies close to the $K\bar{K}$
threshold~\cite{a0}. The final goal of these investigations is 
the extraction of the $a_0/f_0$-mixing amplitude, a quantity which is
believed to shed light on the nature of these
resonances~\cite{Hanhart,Achas79}. As a first stage the reaction
$pp\to d K^+\bar{K^0}$ has been measured at $T_p$=2.65~GeV,
corresponding to an excess energy of $Q$=46~MeV above the
$K^+\bar{K^0}$ threshold, using the ANKE spectrometer~\cite{ank}.
The data, which have been decomposed into partial waves, show that
more that 80\% of the kaons are produced in a relative $s$--wave,
corresponding to the $a_0^+$ channel~\cite{Kleb}. In this paper we
report on the analysis of the reaction $pp\to d\pi^+\eta$, which
was measured in parallel.

\vspace*{-4mm}
\section{The ANKE spectrometer and data analysis}
\vspace*{-4mm}

\begin{figure}[ht]
  \begin{center}
    \leavevmode
    \resizebox{\columnwidth}{!}{\includegraphics[scale=1]{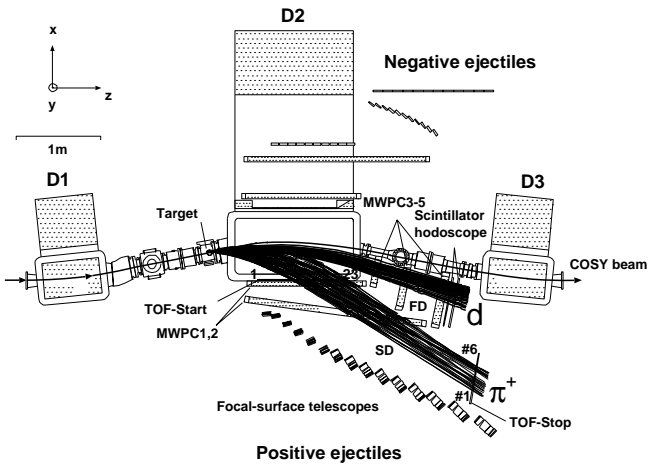}}
    \vspace*{-5mm}
    \caption{Top view of the ANKE spectrometer and simulated tracks of pions and deuterons
             from $pp\to d\pi^+\eta$ events.}
    \label{fig:anke}
  \vspace*{-5mm}
  \end{center}
\end{figure}

Fig.~\ref{fig:anke} shows the layout of ANKE. It
consists of three dipole magnets (D1 -- D3), installed at an
internal target position of COSY.  D1 and D3 deflect the
circulating COSY beam from and back into the nominal orbit. The
C-shaped spectrometer dipole D2 separates forward-going reaction
products from the COSY beam. The angular acceptance of ANKE covers
$|\vartheta_{\mathrm h}| \le 10^{\circ}$ horizontally and
$|\vartheta_{\mathrm v}|\le 3^{\circ}$ vertically for the detected
deuterons ($p_d>1300$~MeV/c), and $|\vartheta_h| \le 12^{\circ}$
and $|\vartheta_v| \le 3.5^{\circ}$ for the pions.  An H$_2$
cluster-jet target~\cite{tar}, placed between D1 and D2, has been
used, providing areal densities of ${\sim} 5 {\times}
10^{14}$\,cm$^{-2}$.  The luminosity has been measured using $pp$
elastic scattering, recorded simultaneously with the $d\pi^+$ data. 
Protons in the angular range $\vartheta= 5.5^\circ - 9^\circ$ have been
selected, since the ANKE acceptance changes smoothly in this
region and the elastic peak is easily distinguished from the
background in the momentum distribution. The average luminosity
during the measurements has been determined as $L=(2.7 {\pm}
0.1_\mathrm{stat} {\pm} 0.7_\mathrm{syst}) {\times}
10^{31}$\,s$^{-1}$\,cm$^{-2}$, corresponding to an integrated
value of $L_\mathrm{int}=3.3$\,pb$^{-1}$ at a proton beam
intensity of up to $\sim 4\times 10^{10}$.

\begin{figure}[ht]
  \begin{center}
    \leavevmode
    \vspace*{-1mm}
    \resizebox{\columnwidth}{!}{\includegraphics[scale=1]{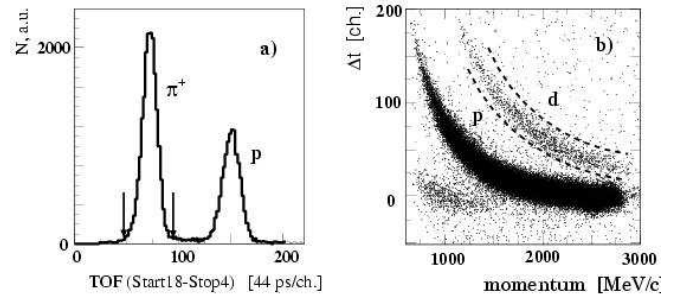}}
    \vspace*{-7mm}
    \caption{a) Time-of-flight of particles detected in TOF start counter 18 and stop counter 4;
             b) time difference $\Delta{t}$ between fast forward-going particles in the FD
and $\pi^+$ mesons as a function of the momentum of the forward
particle. The dashed lines indicate the criteria for deuteron
identification.}
    \label{fig:id}
  \vspace*{-6mm}
  \end{center}
\end{figure}

Two charged particles, $\pi^+$ and $d$, were detected in
coincidence. Their trajectories, simulated with the ANKE-GEANT
program package~\cite{Iza},
are shown in Fig.~\ref{fig:anke}.
Positively charged pions in the momentum range
$p_{\pi}=(600-1100)$~MeV/c were identified in the side detection system
(SD)~\cite{ank, NIM481}, consisting of one layer of 23 start
time-of-flight (TOF) scintillation counters, two multi-wire
proportional chambers (MWPCs) and one layer of 6 counters for
TOF-stop. Pions were selected by a time-of-flight technique
(Fig.~\ref{fig:id}a). The fast deuterons with momenta
$p_{d}=(1300-2800)$~MeV/c (and elastically scattered protons with
$p_p \approx 3400$~MeV/c) hit the forward detection system
(FD)~\cite{ank,FD}, which includes three MWPCs and two layers of
scintillation counters. Two-dimensional distributions $\Delta{t}$
{\em vs.} momenta of forward particles have been used for the
deuteron identification. $\Delta{t}$ was calculated as the time
difference between the detection of a pion in the TOF-stop counter and
a fast forward-going particle in the first layer of the FD scintillators. 
Two distinct bands
corresponding to protons and deuterons are seen in this
distribution (Fig.~\ref{fig:id}b). The selection of deuterons was
achieved by cutting along the deuteron band.

The tracking efficiency for pions in the SD MWPCs was calculated
as the ratio of particles in the proper TOF range with and without
requiring a reconstructed track. The efficiency depends on the SD
stop-counter number (i.e.\ $\pi^+$ momentum) and varies between
$53\%$ and $76\%$.  For the FD MWPCs the efficiency has been
determined for each of the six sensitive planes (two per chamber).
At least two vertical and horizontal planes were demanded for
track reconstruction and for the calculation of the intersection point
with the remaining plane. Each plane has been divided in
20$\times$20 subcells, and the efficiency of each cell has been
calculated from the presence and absence of a hit in the
reconstructed intersection. The average FD MWPC track efficiency
for deuterons is 73$\%$.  The efficiencies of the scintillators
and TOF criteria are larger than 99$\%$~\cite{NIM481}. The
efficiency correction is done on an event-by-event basis.

\section{Results for the reaction $pp \to d\pi^+\eta$}

\begin{figure}[ht]
  \begin{center}
    \leavevmode
    \vspace*{-1mm}
    \resizebox{\columnwidth}{!}{\includegraphics[scale=1]{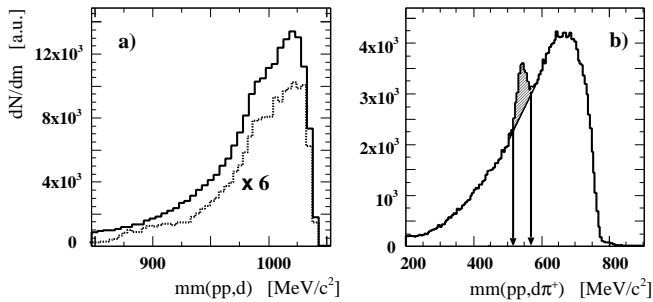}}
    \vspace*{-7mm}
    \caption{Missing mass distributions (a) $mm(pp,d)$,
    (b) $mm(pp,d\pi^+)$ for the reaction $pp\to
    d\pi^+X$. The dotted histogram in $mm(pp,d)$ (scaled by factor 6)
    corresponds to the selected area around the $\eta$ peak ($530-560$~MeV/c$^2$)
    in $mm(pp,d\pi^+)$ (indicated by arrows).}
    \label{fig:pi-et2}
  \vspace*{-4mm}
  \end{center}
\end{figure}

The missing mass distributions $mm(pp,d)$ and $mm(pp,d\pi^+)$ for
the selected $d\pi^+$ pairs are presented in
Fig.~\ref{fig:pi-et2}. In the $(pp,d\pi^+)$ missing mass
distribution a clear peak is observed around
m($\eta$)=547~MeV/c$^2$ with about 6200 events.  The peak sits on
top of a smooth background from multi-pion production $pp\to
d\pi^+(n\pi)$ $(n\ge2)$. After selecting the mass range
$(530-560)$~MeV/c$^2$ around the $\eta$ peak, the missing mass
spectrum $mm(pp,d)$ exhibits a shoulder at 980 MeV/c$^2$
(Fig.~\ref{fig:pi-et2}a, dotted), where the peak from the
$a_0^+$(980) resonance is expected.

\begin{table}[h]
\begin{center}
\begin{tabular}{|c|c|c|c|}
\hline
${\mathrm d}^4\sigma/{\mathrm d}{\Omega_d}{\mathrm d}{\Omega_{\pi^+}}{\mathrm d}p_d{\mathrm d}p_{\pi^+}$ &  \multicolumn{3}{|c|}{ variables, lab. system}\\
${\mathrm {{\mu}b}}/({\mathrm {sr\,GeV/c}})^2$ &$\theta_y$, deg. &$\theta_x$, deg.& $p$, GeV/c  \\
\hline
&& & \\
 &  \multicolumn{3}{|c|}{deuteron variables} \\
$71\pm 6_{\mathrm{stat}}\pm 20_{\mathrm{sys}}$ & $(-3^{\circ},+3^{\circ})$ & $(-3.5^{\circ},+3.5^{\circ})$ & $(1.4,1.6)$ \\
 &  \multicolumn{3}{|c|}{pion variables} \\
 & $(-4^{\circ},+4^{\circ})$ & $(-11^{\circ},-3^{\circ})$ & $(0.65,0.95)$ \\
&& & \\
\hline
&& & \\
 &  \multicolumn{3}{|c|}{deuteron variables} \\
$30\pm 4_{\mathrm{stat}}\pm 9_{\mathrm{sys}}$ & $(-3^{\circ},+3^{\circ})$ & $(-6^{\circ},-2^{\circ})$ & $(1.8,2.7)$ \\
 &  \multicolumn{3}{|c|}{pion variables} \\
 & $(-4^{\circ},+4^{\circ})$ & $(-11^{\circ},-3^{\circ})$ & $(0.65,0.95)$ \\
&& & \\
\hline
\end{tabular}
\caption{Differential production cross sections for the reaction
$pp \to d\pi^+\eta$. In both regions of phase space $a_0^+$ and
non-resonant $\pi^+\eta$
  productions contribute. For the momentum range
  $p_d=(1.4-1.6)$~GeV/c the non-resonant $\pi^+\eta$  production should be
  dominant, because this momentum range corresponds to low masses of
  the $\pi^+\eta$ system, where the $a_0^+$ production is suppressed.}
\label{table:cross}
\end{center}
\end{table}

Table~\ref{table:cross} presents the differential cross sections
for $pp \to d\pi^+\eta$ for two regions of phase space where the
acceptance of ANKE is essentially $100\%$ for this reaction. Six
variables in the lab.\ system have been chosen to describe these
rectangular areas: the vertical ($\theta_y$) and horizontal
($\theta_x$) angles and momenta of the two detected particles,
deuteron and $\pi^+$. These angles are defined as
$\tan(\theta_y)=p_y/p_z$, $\tan(\theta_x)=p_x/p_z$.

Unfortunately, due to the limited phase-space coverage of ANKE, a
partial wave decomposition of the type performed in
Ref.~\cite{Kleb} is not possible in this case. Thus, in order to
interpret the data in terms of $a_0^+$ and non--resonant $\pi^+
\eta$ production, we need to employ a model. This investigation
will be described in the next section.

\section{Interpretation of the results}

The $d\pi^+\eta$ final state can be produced \textit{via} the
$a_0^+$ resonance, $pp\to da_0^+\to d\pi^+\eta$ (resonant
production), or through the direct reaction $pp\to d\pi^+\eta$
(non-resonant production). The production mechanism for the
$a_0^+$ has been studied theoretically in Refs.~\cite{Gri521,
Mull, Oset, Brat, Kon03}. According to Refs.~\cite{Gri521, Mull},
the cross section for $a_0^+$ production at $T$=2.65~GeV is
expected to be $\sim 1 \mu$b while, for the non-resonant
$\pi^+\eta$ production, different predictions exist, ranging from
$(0.6-1.4)~\mu$b~\cite{Tarasov} and $(1.6-3.3)\mu$b~\cite{Gri00}
up to one order of magnitude more~\cite{Mull}.

\begin{figure}[ht]
  \begin{center}
    \leavevmode
    \resizebox{\columnwidth}{!}{\includegraphics[scale=1]{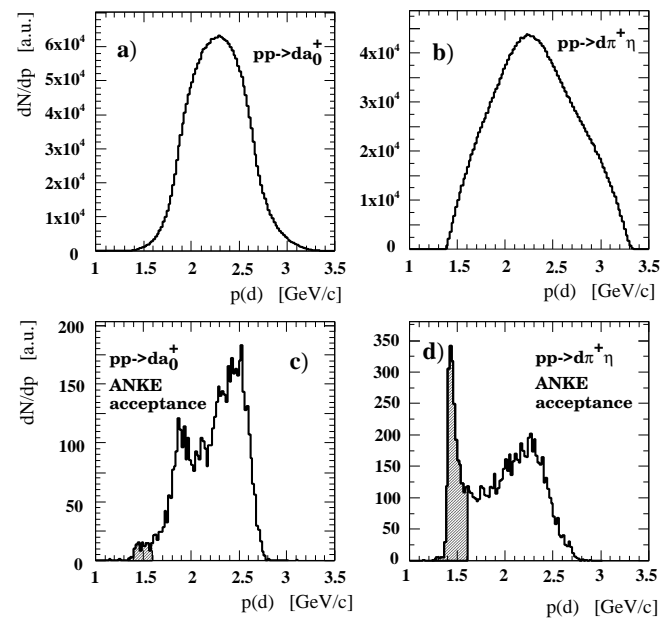}}
    \vspace*{-7mm}
    \caption{Simulated acceptance for the deuteron momentum for
    the resonant (a,c) and the non-resonant (b,d) $\pi^+\eta$
    production.  Upper row: the initial model distributions~\cite{Gri04,Tarasov2,Gri00,Kon03}; lower row:
    momentum distributions within the ANKE acceptance.}
    \label{fig:pd}
  \vspace*{-6mm}
  \end{center}
\end{figure}

\begin{figure}[ht]
  \begin{center}
    \leavevmode
    \resizebox{\columnwidth}{!}{\includegraphics[scale=1]{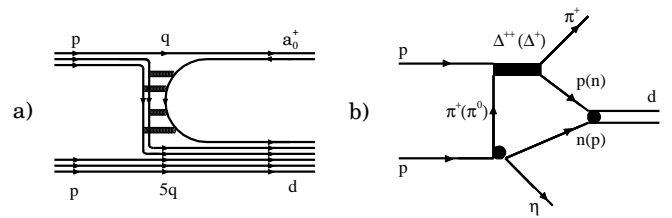}}
    \vspace*{-7mm}
    \caption{Diagrams describing the resonant
       $\pi^+\eta$ production within the QGSM~\cite{Gri04} (a) and the non-resonant
        \textit{via} $N^*$ and $\Delta$ excitations~\cite{Tarasov2,Gri00,Kon03} (b).}
    \label{fig:mod}
  \vspace*{-6mm}
  \end{center}
\end{figure}

The differential cross sections from Table~\ref{table:cross}
contain contributions from both the resonant and the non-resonant
reactions. We have performed a model-dependent analysis using the
calculated momentum distributions of the produced deuterons (see
Figs.~\ref{fig:pd}a,b). The calculations are based on models
describing the resonant process within the Quark-Gluon Strings
Model (QGSM) (Fig.~\ref{fig:mod}a)~\cite{Gri04} and the
non-resonant (Fig.~\ref{fig:mod}b) production \textit{via} $N^*$-
and $\Delta$-resonance excitation~\cite{Tarasov2,Gri00,Kon03}. The
momentum distribution for the resonant reaction is narrower,
because low $(\pi^+\eta)$ masses are suppressed for $a_0^+$
production.  The momentum distributions within the ANKE acceptance
are shown in Figs.~\ref{fig:pd}c,d.

\begin{figure}[ht]
  \begin{center}
    \leavevmode
    \resizebox{\columnwidth}{!}{\includegraphics[scale=1]{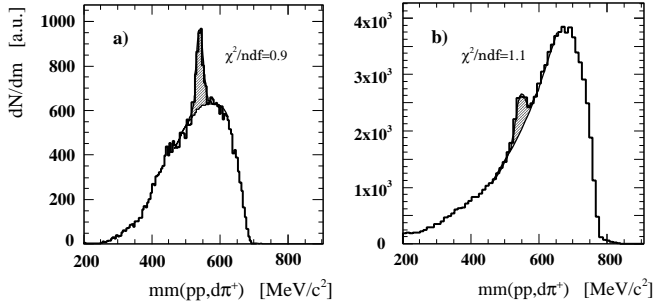}}
    \vspace*{-7mm}
    \caption{Missing mass distribution $mm(pp,d\pi^+)$ for
    the momentum ranges $p_d=(1.4-1.6)$~GeV/c (a) and $p_d=(1.6-2.8)$~GeV/c (b).}
    \label{fig:pdcut3}
  \vspace*{-6mm}
  \end{center}
\end{figure}

Within the models the non-resonant background has a negligible
$\pi \eta$ $s$--wave contribution and thus does not interfere with
the resonant amplitude. As a consequence, in the range
$p_d=(1.4-1.6)$~GeV/c, the contribution of the non-resonant
process dominates whereas the resonant part is negligibly small
(see shaded areas in Figs.~\ref{fig:pd}c,d).  Thus, the cross
section of the non-resonant contribution in this momentum range
can be defined by fitting the $\eta$ peak in $mm(pp,d\pi^+)$
(Fig.~\ref{fig:pdcut3}a) and extracting the number of $d\pi^+\eta$ events.  
The missing mass spectra have been
described by the sum of a Gaussian distribution and a
$4^\mathrm{th}$ order polynomial.

In the momentum range $p_d=(1.6-2.8)$~GeV/c, where both the
resonant and the non-resonant reactions contribute, it is possible
to calculate the number of events from the non-resonant production
taking into account the different acceptances. Then the number of
$a_0^+$ events is the difference between the total number of
events under the $\eta$ peak (Fig.~\ref{fig:pdcut3}b) and the
calculated amount of non-resonant $\pi^+\eta$ events.

Including all corrections, total cross sections
$\sigma_{\mathrm{a_0}}=(1.1\pm 0.3_{\mathrm{stat}}\pm
0.7_{\mathrm{sys}})~\mathrm{\mu{b}}$ for the $a_0^+$ production
and $\sigma_{\mathrm{n.r.}}=(3.5\pm 0.3_{\mathrm{stat}}\pm
1.0_{\mathrm{sys}})~\mathrm{\mu{b}}$ for the non-resonant
$\pi^+\eta$ production have been obtained. It is obvious, that
these numbers could change if different assumptions were made on
the partial wave decomposition of the (non--)resonant contributions.
However, we here refrain from investigating further the
uncertainty induced by the model assumptions.

\begin{figure}[ht]
  \begin{center}
    \leavevmode
    \vspace*{-1mm}
    \resizebox{\columnwidth}{!}{\includegraphics[scale=1]{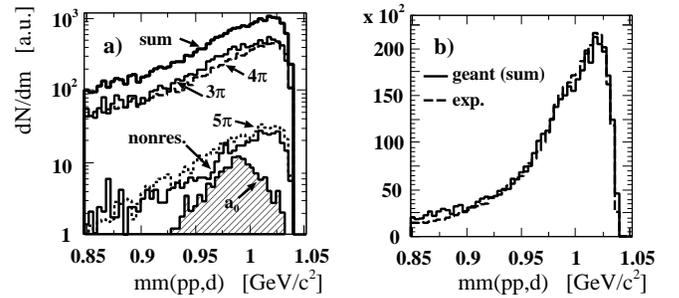}}
    \vspace*{-7mm}
    \caption{a) GEANT simulations for multi-pion background ($pp\to
    d\pi^+(n\pi)$ $(n\ge2)$), and for resonant and non-resonant
    production of the $d\pi^+\eta$ final state in the ANKE
 acceptance. b) comparison between the simulated missing mass
    distribution $mm(pp,d)$ and experimental data.}
    \label{fig:multi}
  \vspace*{-6mm}
  \end{center}
\end{figure}

Figure~\ref{fig:multi}a presents the results of GEANT simulations
of the ANKE acceptance for multipion background ($pp\to
d\pi^+(n\pi)$ $(n\ge2)$), and for resonant and non-resonant
production of the $d\pi^+\eta$ final state. The number of initial
events for each reaction is proportional to the known cross
sections for multipion production~\cite{Bug} and the values of the
total cross sections for the resonant and non-resonant $\pi^+\eta$
production given above. The shape of the simulated missing mass
distribution $mm(pp,d)$ including all channels is in good
agreement with the experimental data (Fig.~\ref{fig:multi}b).

Data for the second $a_0^+$ decay channel, $a_0^+\to
K^+\bar{K^0}$, have been obtained at ANKE in the reaction $pp \to
da_0^+\to dK^+\bar{K^0}$, simultaneously with the $\pi^+\eta$
data. The measured total cross section is $\sigma(pp \to
dK^+\bar{K^0})=(38\pm 2_{\mathrm{stat}}\pm 14_{\mathrm{sys}})$~nb
and the contribution of the ${(K^+\bar{K^0})}_{L=0}$ channel is
$\sim 83\%$~\cite{Kleb}.  Assuming that $s$--wave $(K^+\bar{K^0})$
production proceeds fully \textit{via} the $a_0^+$(980) resonance,
in accordance with the predictions of the models discussed above,
the ratio of the total cross sections is $R=\sigma(a_0^+\to
{(K^+\bar{K^0})}_{L=0})/\sigma(a_0^+\to \pi^+\eta)=0.029\pm
0.008_\mathrm{stat}\pm 0.009_\mathrm{sys}$.

\begin{figure}[ht]
  \begin{center}
    \leavevmode
    \vspace*{4mm}
    \resizebox{\columnwidth}{6cm}{\includegraphics[scale=1]{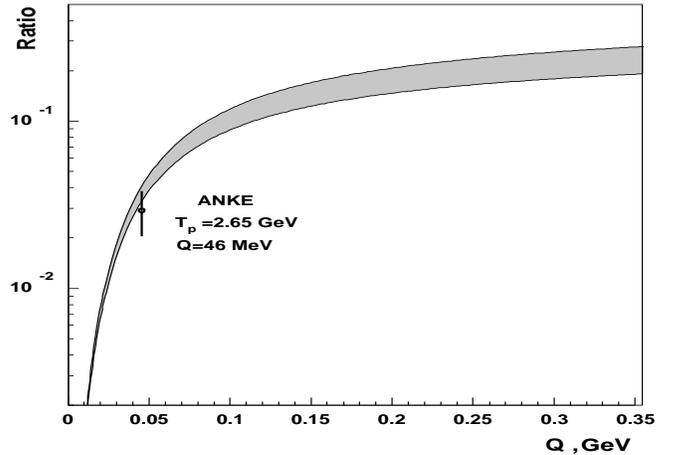}}
  \vspace*{-6mm}
    \caption{Ratio of the total cross sections $\sigma(a_0^+\to
    K^+\bar{K^0})/\sigma(a_0^+\to \pi^+\eta)$ as function of $Q$.
    The error bar shows the statistical error only.}
    \label{fig:br}
  \end{center}
\end{figure}

For measurements at an excess energy $Q>>\Gamma_{a_0}\sim
(50-100)$~MeV, the ratio of the total cross sections
$R=\sigma(a_0^+\to K^+\bar{K^0})/\sigma(a_0^+\to \pi^+\eta)$
should not depend upon $Q$.  However, when $Q$ is of the order of
$\Gamma_{a_0}$, the contribution from the $a_0^+\to K^+\bar{K^0}$
decay channel decreases more strongly due to the proximity of the
$K^+\bar{K^0}$ threshold and the consequently limited available
phase space. The calculated ratio of the two cross sections as a
function of $Q$ is presented in Fig.~\ref{fig:br}. The calculation
is normalised to $R((K\bar{K})/(\pi\eta))=0.23\pm 0.05$, which has
been measured at Crystal Barrel in the reaction $p\bar{p}\to
a_0\pi$ at $Q=768$~MeV~\cite{Abe98}. The parameters of the
Flatt\'e distribution~\cite{Flatte} $m_0=999\pm 2$~MeV/c$^2$,
$g_{\pi\eta}=(324\pm 15)$~MeV and
$r=g_{KK}^2/g_{\pi\eta}^2=1.03\pm 0.14$ are also taken from
Ref.~\cite{Abe98}. The errors of the Flatt\'e parameters define
the range of possible $R$ values (shaded area in
Fig.~\ref{fig:br}).  Our result is in agreement with the
calculated value.

\section{Conclusions}

To summarize, data on the reaction $pp\to d\pi^+\eta$ at
$T_p=2.65$~GeV are presented. Differential cross sections for
limited regions of phase space, corresponding to forward emission
of the  $d$ and $\pi^+$ in the laboratory system, have been
extracted. Dynamical models for the reaction were employed to
obtain the total cross sections for the resonant $\pi^+\eta$
production \textit{via} the $a_0^+$(980) and for the non-resonant
$\pi^+\eta$ channel. For $a_0^+$ production the value for the
total cross section $pp \to da_0^+ \to d\pi^+\eta$ is in agreement
with theoretical predictions based on the value for the $pp\to
d(K^+ \bar{K^0})_{L=0}$ reaction recently measured at
ANKE~\cite{Kleb} and the branching ratio $BR(K\bar{K}/\pi\eta)$
from the literature.

Our results, together with those from ANKE on the decay channel
$a_0^+\to K^+\bar{K^0}$, are the first evidence for $a_0^+$
production in $pp$ collisions, obtained in a simultaneous
exclusive measurement of both $a_0^+$ decay channels.

\vspace*{4mm} We are grateful to C.~Wilkin for critical
discussions and a careful reading of the manuscript. We would
like to thank to W.~Borgs, C.~Schneider and the IKP technicians for the
support during the beamtime. One of the authors (P.F.)
acknowledges support by the COSY-FFE program (grant FFE-41520733).
The work at ANKE has partially been supported by: BMBF (grants
WTZ-RUS-686-99, 211-00, 691-01, GEO-001-99, POL-007-99, 015-01),
DFG (436 RUS 113/444, 630, 733,787), DFG-RFBR 03-02-04013,
02-02-04001(436 RUS 113/652), Polish State Committee for
Scientific Research (2 P03B 101 19), Russian Academy of Science
(RFBR 06518, 02-02-16349), ISTC (1861, 1966).

\end{document}